\documentclass[12pt,preprint]{aastex}
\usepackage{amsmath}
\usepackage{graphicx}
\usepackage{epsfig}

\usepackage{url}
\usepackage{indentfirst}
\begin{document}
\title{X-ray/GeV emissions from Crab-like pulsars in LMC }
\author{Takata, J.\altaffilmark{1} \and Cheng, K. S.\altaffilmark{2}}
\email{takata@hust.edu.cn, hrspksc@hku.hk}
\altaffiltext{1}{School of physics, Huazhong University of Science and Technology, Wuhan 430074, China}
\altaffiltext{2}{Department of Physics, The University of Hong Kong, Pokfulam Road, Hong Kong}
\begin{abstract}
  We discuss X-ray and gamma-ray emissions from Crab-like pulsars, PSRs~J0537-6910 and~J0540-6919, in Large Magellanic Cloud.
  Fermi-LAT observations have resolved the gamma-ray emissions from these two pulsars and found the pulsed
  emissions from PSR~J0540-6919. The total pulsed radiation in the X-ray/gamma-ray energy bands of  PSR~J0540-6919 is observed
  with the efficiency $\eta_{J0540}\sim 0.06$ (in 4$\pi$ sr), which is about
  a factor of ten larger than $\eta_{Crab}\sim 0.006$ of the Crab pulsar.  Although PSR~J0537-6910 has  the highest spin-down
  power among currently known pulsars, the
  efficiency of the observed X-ray emissions is   about two orders of magnitude smaller than that of PSR~J0540-6919.
  This paper mainly  discusses what causes  the  difference in the radiation efficiencies of these three energetic Crab-like pulsars. We discuss electron/positron acceleration and high-energy emission processes within
  the outer gap model. By solving
  the outer gap structure with the dipole magnetic field, we show that the radiation efficiency decreases as the inclination
  angle between the magnetic axis and the rotation axis increases.  To explain the difference in
  the pulse profile and in the radiation efficiency, our model suggests that PSR~J0540-6919 has
  an inclination angle much smaller than the that of Crab pulsar (here we assume the inclination angles
  of both pulsars are $\alpha<90^{\circ}$). On the other hand, we speculate that the difference in the radiation efficiencies
  between PSRs~J0537-6910 and J0549-6919 is  mainly caused by the difference in the Earth viewing angle, and that
  we see PSR~J0537-6910 with an Earth viewing angle $\zeta>>90^{\circ}$ (or $<<90^{\circ}$) measured from the
  spin axis, while we see PSR~J0540-6919 with  $\zeta\sim 90^{\circ}$.
  
\end{abstract}

\section{Introduction}
\begin{table}
  \begin{tabular}{ccccccc}
    \hline
    PSRs & $P_s$ & $L_{sd,38}$ & $B_{lc,6}$  & $d_{kpc}$ &  $\eta_x$ & $\eta_{\gamma}$ \\
    \hline
    J0537-6910$^{a}$ & 0.016& 4.9 & 2.07 & 50 & $3\times 10^{-4}$ & - \\
    Crab$^{b}$ & 0.033& 4.5& 0.96 & 2&  $5\times 10^{-3}$  & $10^{-3}$  \\
    J0540-6919$^{c}$ & 0.05 & 1.5& 0.36 & 50  &0.024 &0.038\\
    J1813-1749 & 0.045& 0.56 & 0.25 &4.7& - & - \\
    J1400-6325$^{d}$ & 0.03 & 0.51 & 0.35 & 7&  $10^{-3}$ & $<1.5\times 10^{-3}$  \\
\hline
  \end{tabular} 
  \caption{The five most energetic pulsars: From the left to the right columns, pulsar name (PSR), rotation period ($P_s$)  in units of second,  spin down age $(L_{sd,38}$) in units of $10^{38}{\rm erg~s^{-1}}$,
    the magnetic field strength at the light cylinder ($B_{lc,6}$) in units of $10^{6}$G, distance to the source
    $(d_{kpc})$ in units of kpc,  X-ray efficiency and gamma-ray efficiency (in $4\pi$ radian).  $a$; the X-ray efficiency in 2-10keV
    energy bands (Mineo et al. 2004). $b$; the X-ray efficiency in 0.3-10keV and gamma-ray efficiency above 100MeV
    (Abdo eta l. 2013). $c$; the X-ray efficiency calculated with  ``absorbed'' fluxes in 2-10keV and 20-100keV (Campana et al. 2008) and gamma-ray efficiency above 100MeV (Ackermann et al. 2015). $d$; the efficiency in 20-100keV,
    including pulsar and PWN  and upper limit of gamma-ray efficiency (Renaud et al. 2010 and reference therein).}
\end{table}

PSRs~J0537-6910 and J0540-6919 are energetic young pulsars in the Large Magellanic
Cloud (hereafter LMC), and they were discovered by the X-ray observations (Seward et al. 1984; Marshall et al. 1998).
The spin-down powers of PSRs~J0537-6910 and J0540-6919 are $L_{sd}\sim 5\times 10^{38}{\rm erg~s^{-1}}$
and $\sim 1.5\times 10^{38}{\rm erg~s^{-1}}$, respectively, which are similar
to $L_{sd}\sim 4.5\times 10^{38}\rm {erg~s^{-1}}$ of the Crab pulsar. Among currently known pulsars, these three,
PSR~J0537-6910, Crab and J0540-6919, have the top three highest spin-down power (see Table~1). In this paper,
``Crab-like pulsars'' is used to refer to all three.
 The Fermi Large Area Telescope (hereafter Fermi-LAT) resolved the gamma-ray emissions from the two 
Crab-like  pulsars in the LMC,  and furthermore detected the pulsed emissions from PSR~J0540-6919 (Ackermann et al. 2015). 

PSR~J0540-6919 (spin period $P_s=0.05$s) is known as the ``Crab-twin'', because
not only the spin-down parameters but also the properties of the pulsed
emissions in multi-wavelength bands
are similar to those of the Crab pulsar. First,  Fermi-LAT found 
that the ratio of X-ray luminosity and  $>0.1$GeV gamma-ray luminosity
is $L_{X}/L_{\gamma}\sim 1$, which is similar to $L_{X}/L_{\gamma}\sim 5$ for the Crab pulsar (Abdo et al. 2010).
This feature is clearly distinct from $L_{X}/L_{\gamma}<10^{-3}-10^{-4}$ of the others (Abdo et al. 2013
for the Fermi-LAT pulsar catalog).  Second, the pulse peaks in different wavelength bands are
all in  phase, just like the
pulse profiles of the Crab pulsar.  Furthermore, PSR~J0540-6919 emits the 
giant radio pulses that appear  at the positions of the pulse peaks in higher-energy
bands (Johnston et al 2004). This property is also  the same as for 
the Crab pulsar (Shearer et al. 2003). It is likely  that there  three features
  represent the nature of the  pulsars
 with $L_{sd}>10^{38}{\rm  erg~s^{-1}}$.

 While  the Crab-like pulsars are similar in
 their spin-down properties,
there are several remarkable differences in
the observed radiation: (1) the pulse shape, and (2)
the radiation efficiency, which is defined as 
the ratio of the radiation luminosity to the spin-down
power $\eta\equiv L_{rad}/L_{sd}$. PSR~J0540-6919 shows a broad pulse profile with a small dip at
the center (Campana et al. 2008 for the  X-ray pulse and Gradari et al. 2011 for the optical pulse), while
the Crab pulsar shows a sharp double-peak structure
with the phase separation of $\delta\phi\sim 0.4$ (Abdo et al. 2010).  The integrated luminosity of the
pulsed X-ray/gamma-ray emissions from PSR~J0540-6919
is $L_{rad}\sim 10^{37}(d/50{\rm kpc})^2$erg/s (in $4\pi$ sr), which is about a factor of 3 larger
than that of the Crab pulsar, $L_{rad}\sim 3\times 10^{36}(d/2{\rm kpc})^2$erg/s. As a result, the radiation efficiency  $\eta_{J0540}\sim 0.06$  of
 PSR~J0540-6019 is a factor of ten larger than that of the Crab pulsar, $\eta_{Crab}\sim 0.006$.
The Fermi-LAT confirmed  that
 the luminosity of the non-thermal radiation from a pulsar tends to
 increase as  $L_{\gamma}\propto L^{1/2}_{sd}$ (Abdo et al. 2013), which yields $\eta\propto L_{sd}^{-1/2}$.
 This empirical relation cannot explain the ratio $\eta_{J0540}/\eta_{Crab}\sim 10$.
 For PSR~J0537-6910, the Fermi-LAT did not detect the pulsed emissions, and
 it measured the  spectrum fitted by a power-law function, suggesting  the emissions originate  from the
 pulsar wind and/or a supernova remnant. The pulsed X-ray emissions from PSR~J0537-6910 were observed
  to be $F_{X}\sim 5\times 10^{-13}{\rm erg~cm^{-2}~s^{-1}}$ in 2-10keV (Mineo et al. 2004), indicating
 the efficiency is $\eta_{J0537}\le 10^{-3}$, which is much lower than those of the Crab and PSR~J0540-6919.
 Since the distance to the Crab ($d\sim 2$kpc), and
 the  LMC ($d\sim 50$kpc) are well determined, the uncertainty of the efficiency due to that is the distance should be small.
 The observed emission properties of the three
   energetic  pulsars pose  a challenge theoretically to see whether a  unique model can explain all these systems.

  Electron/positron  acceleration and high-energy emission processes
   in  the pulsar magnetosphere have recently been discussed within  the framework of the 
   slot gap model (Harding et al. 2008; Harding \& Kalapotharakos 2015), outer gap model (Cheng et al. 2000, Hirotani 2015, Takata
   et al. 2016),  current sheet of the force-free magnetosphere model (Spitkovsky 2006; Bai \& Spitkovsky 2010), and
   pulsar wind model (Aharonian et al. 2012). In this paper,
   we will discuss the high-energy emission process
within the framework of the outer gap accelerator model. For the Crab-like pulsars,
the outer gap model has predicted
that most of  $>$GeV photons from the outer gap are converted into pairs by the pair-creation process
and cannot escape from the light cylinder (see section~\ref{model}, Cheng et al. 2000; Takata \& Chang 2007;
Tang et al. 2008). Synchrotron radiation and the inverse-Compton process of the secondary pairs
can produce the observed emissions in the optical to TeV energy bands.  
The outer gap model predicts that 
the shape of the pulse profile is sensitive to the viewing angle and magnetic  inclination angle measured from the spin axis.
In the Fermi-LAT pulsar catalog (Abdo et al. 2013), $\sim 75\%$ of the sources show  a double-peak structure in the
  pulse profile and $\sim 40\%$ show a wide phase (0.4$\sim$ 0.6) separation between the two peaks.
The outer gap model explains the widely  separated two peaks by  assuming a larger magnetic inclination angle and a larger Earth
viewing angle  (Takata et al. 2011; Watters \& Romani 2011).
On the other hand, Takata \& Chang (2007) explain the pulse profile
of PSR~J0540-6919 by  a smaller inclination angle $\alpha\sim 30^{\circ}$ and a larger viewing angle
$\zeta\sim 90^{\circ}$. The observed geometry of the pulsar wind tori also
 suggests the viewing angle $\zeta\sim 90^{\circ}$ for PSR~J0540-6919 (Ng \& Romani 2004, 2008).

Previous studies of PSR~J0540-6919 (Zhang \& Cheng 2000; Takata \& Chang 2007) mainly discussed the optical/X-ray emissions, since only the upper limit of
the GeV flux had been reported before the launch of the Fermi. Hence, 
it is not obvious why the efficiencies of the observed radiations among 
the Crab-like pulsars are so different. In this paper, therefore, we will
revisit the  non-thermal emission process of the Crab-like pulsars with the 
outer gap model.  In section~2, we will describe our theoretical model for  the Crab-like pulsars.
In section~3, we present our result
of the fitting spectrum for PSR~J0540-6919  and discuss
 the differences  between the Crab and this pulsar. 
In section~4, we will discuss the emissions from  PSR~J0537-6910.

\section{Theoretical Model}
\label{model}
We apply the calculation method developed in Takata et al. (2016), which
solves the outer gap structure in the three-dimensional space  with a  rotating  dipole magnetic field. They obtained
the structure of the accelerating electric field and gap currents by solving the Poisson equation, and the
continuity equations for the electrons and positrons and the  pair-creation process. For the outer gap model, some electrons/positrons, which migrate
along the magnetic field lines, should enter the outer gap from the gap boundaries and they initiate
the gamma-ray radiation and subsequent pair-creation cascade processes.
Takata et al. (2016)  solved the pair-creation cascade inside the outer gap by assuming the number of  the electron/positron  injections at the
gap boundaries, which is the crucial factor for  control of the outer gap structure.
We refer Takata et al (2016) for detailed calculations.

Takata et al. (2016)  assumed that the outer gap structure
is variable in time,  rather than stationary, because of the time-dependent injection
of the electrons/positrons at the gap boundaries. The model argued that
 the observed gamma-ray spectrum is a superposition of the emissions
 from different  stationary gap structures with different injection rates at the gap
 boundaries. This dynamic model provides a better fit for  the spectra of the Fermi-LAT pulsars.
 As we will argue later (see section~\ref{discuss}), our model suggests that
 the observed emissions from the Crab-like pulsars are
 not from primary electrons/positrons accelerated  in the outer gap, but from
 the  secondary pairs created outside the gap, and therefore 
 the dynamic behavior of the outer gap may be less important
 in  explaining  the pulsar's observed GeV spectra. 

 In our calculations, the model parameters that determine the gap dynamics
 are the inclination angle of the magnetic axis, the surface temperature of the neutron star, and the number
 of the electrons and positrons that enter the outer gap from the gap boundaries.
 We assume the magnetic inclination angle $\alpha$ less than $90^{\circ}$
 measured from the spin axis.  For the rotator with a  $\alpha<90^{\circ}$,
 the positrons and electrons can enter the gap from outside along the magnetic field lines by crossing  the inner
 boundary (star side)  and outer boundary (light cylinder side), respectively.
 We assume that the rate of the particle
 injection is  constant over the inner and the outer boundaries.  
 We parameterize the injection current in units of the Goldreich-Julian value  and denote
 $j_{in}$ and $j_{out}$ as the normalized injection rates at the gap inner and outer boundaries, respectively.
 Our local model has to treat the injection currents ($j_{in},~j_{out}$) as the  model fitting parameters.
 Takata et al. (2016) discussed  the origin of 
 the electrons/positrons that enter the outer gap at the gap boundaries.

 In the outer gap, the positrons and electrons crossing the gap boundaries
 initiate the gamma-ray emission and
 a  subsequent pair-creation cascade. The electrons/positrons are accelerated by the electric field parallel
 to the magnetic field and emit the gamma-rays via the
 curvature radiation process and/or the inverse-Compton scattering process.
 The emitted gamma-rays may be converted into pairs, by the pair-creation process,  with
 the surface X-rays. The new pairs created in the gap are accelerated by the electric field and
 emit the curvature photons.   The pairs created outside the gap lose their energy
 via the synchrotron radiation and the inverse-Compton scattering process.
 We denote as   ``primary'' pairs as the electrons/positrons accelerated inside the gap,  and as ``secondary''
 pairs those  produced outside the gap.  Since no measurements 
 on the surface temperature of the Crab-like pulsars have been made,
 we assume $T_s=10^6$K as  the temperature of the entire stellar
 surface.
 
The main difference from the calculation in  Takata et al. (2016) is that the current
 model of the Crab-like pulsars takes into account  emission
 from the pairs created outside the outer gap.
 The X-rays produced by  the synchrotron radiation of the secondary pairs become
 the target soft-photon field for the photon-photon pair-creation process occurring  outside the gap.
 One important difference in the circumstellar   conditions
 between the Crab-like pulsars  and other Fermi-LAT pulsars is the mean-free
 path of the pair-creation process between
 a $>1$GeV photon and a background soft photon produced by the secondary pairs. The optical
 depth inside the light cylinder may be written down as
 \begin{equation}
   \tau_{p}(r)\sim rn_X\sigma_{\gamma\gamma}
   \sim 1\left(\frac{L_{X}}{10^{35}{\rm erg~s^{-1}}}\right)
     \left(\frac{r}{\varpi_{lc}}\right)^{-1}\left(\frac{P_s}{0.05{\rm s}}\right)^{-1}
     \left(\frac{E_X}{0.1{\rm keV}}\right)^{-1},
   \end{equation}
 where  $n_X$ is the number density of the soft photons, $\sigma_{\gamma\gamma}\sim 0.2\sigma_T$ is the cross-section,
 $L_{X}$ is the X-ray luminosity, $E_{X}$ is the energy of the soft photon, and  $\varpi_{lc}=cP_s/2\pi$
 is the light cylinder radius. The optical depth of the Crab-like pulsars is usually
 larger than unity with $L_{X}>10^{35}{\rm erg~s^{-1}}$. For the Crab-like pulsars, therefore,
 most of the primary gamma-rays  with an energy $>1$GeV are absorbed by the pair-creation process and
 the secondary particles   will emit the X-rays via  synchrotron emission. This explains the ratio
 of X-ray and gamma-ray luminosity
 $L_{X}/L_{GeV}\sim 1$  for the Crab-like  pulsars, where $L_{\rm {GeV}}$ is the {\it apparent}
 luminosity from the magnetosphere.
 For other Fermi-LAT pulsars, the mean-free path is of order  $\tau_{p}\sim 10^{-3}$ with $L_{X}\sim 10^{32-33}{\rm erg~s^{-1}}$,
 and results in  $L_{X}/L_{GeV}\sim 10^{-3}$. 

 To calculate the emission from the pairs produced outside the outer gap, we trace the propagation
 of the $>$GeV photons and the pair-creation rates on the trajectory. We calculate the pair-creation
 mean-free path by assuming  the number density  of the soft-photons inferred from the observations; that is,
 \begin{equation}
\frac{dN_s(r)}{dE}=\left(\frac{d}{r}\right)^2\frac{dN_{obs}}{dE}, 
\label{xnumber}
 \end{equation}
 where $dN_{obs}/dE$ is the observed spectrum in the optical  to  hard X-ray energy bands and  $d=50$kpc is the
 distance to the LMC.   The pitch angle, $\theta_p$,
 of the newborn pairs produced outside the outer gap is calculated from
 \begin{equation}
   \cos\theta_p=\mbox{\boldmath$b$}(\mbox{\boldmath$r$})
   \cdot\mbox{\boldmath $n$}_{\gamma}(\mbox{\boldmath$r$}_0),
   \label{pitch}
 \end{equation}
 where $\mbox{\boldmath $b$}$ is the unit vector of the magnetic field, and
 $\mbox{\boldmath$n$}_{\gamma}$ is the propagation direction of the gamma-rays, and 
 $\mbox{\boldmath$r$}$ and $\mbox{\boldmath$r$}_0$ represent the positions of
 the pair-creation and the radiation, respectively. The emission direction is calculated
 from
 $\mbox{\boldmath$n$}_{\gamma}(\mbox{\boldmath$r$}_0)=\beta_0\mbox{\boldmath$b$}+\mbox{\boldmath$\beta$}_{co}$,
 where $\mbox{\boldmath$\beta$}_{co}$ is the co-rotation velocity,
 and $\beta_0$ is calculated from $|\mathbf{n}_{\gamma}|=1$. In the calculation, there is  an  uncertainty in
 the collision angle between the gamma-ray and magnetospheric soft photons. Since the latter 
 are emitted by the secondary  pairs,  which has a pitch angle $\theta_p$, we may assume a collision
 angle of  $\theta_{c}\sim2\theta_p$.

 Grand based Cherenkov telescopes have observed
 the pulsed emissions  up to $\sim $ 1TeV from the Crab pulsar (Abdo et al. 2010; Aleksi$\rm{\acute{c}}$ et al. 2011, 2012, 2014; Aliu et al. 2008, 2011). The emissions
 between 10GeV and 1TeV are well fitted by a single  power-law function. The standard curvature
 radiation process cannot easily explain the emissions above 100GeV from the Crab pulsar,
  which suggests the inverse-Compton scattering  process inside  the magnetosphere  (Aleksi$\rm{\acute{c}}$ et al. 2011; Harding and Kalapotharakos 2015)
 or at the pulsar wind region (Aharonian et al. 2012).  Within the framework of the outer gap scenario,
 the $>100$GeV emissions of the Crab pulsar are  explained by the emission process of
 TeV  primary pairs and/or secondary pairs that  were produced by the pair-creation process
 of the TeV  photons from the inverse-Compton scattering process of the primary pairs. If the infrared (IR) photons from
 the secondary pairs  enter the outer gap, they are up-scattered  by $\sim 1$TeV electrons/positrons
 whose Lorentz factor is $\Gamma\sim 3\times 10^{7}$, and become $\sim 10$TeV gamma-rays. 
 Most of  the TeV gamma-rays from the outer gap are absorbed by the soft photons outside the gap, and
 create $\sim 1$TeV electrons/positrons. The TeV secondary pairs also emit photons via 
 synchrotron radiation and inverse-Compton scattering processes. Furthermore, the high-energy
 secondary photons also targets  for the pair-creation.  In this paper,
 we also examine the pair-creation cascade outside the outer gap which is  initiated
 by the TeV gamma-rays from the outer gap, and we will discuss its contribution
 to the observed emissions of  PSR~J0540-6919.  Since the IR photons are produced above the outer gap, we assume that
  they irradiate the outer gap at around the upper boundary, say $\sim 10$\% of the gap thickness,
 with the number density estimated from equation~(\ref{xnumber}).
 Since  the emission direction of the IR will be  related to  the pitch angle in equation~(\ref{pitch}),
 we may roughly  estimate  the collision angle of the inverse-Compton
 scattering as  $\cos\theta_{IR}\sim \mbox{\boldmath$b$}\cdot \mbox{\boldmath$n$}_{\gamma}$
 at the emission point.

\section{Results}
\subsection{Multi-wavelength spectrum}

\begin{figure}
  \centering
  \includegraphics[width=0.9\textwidth]{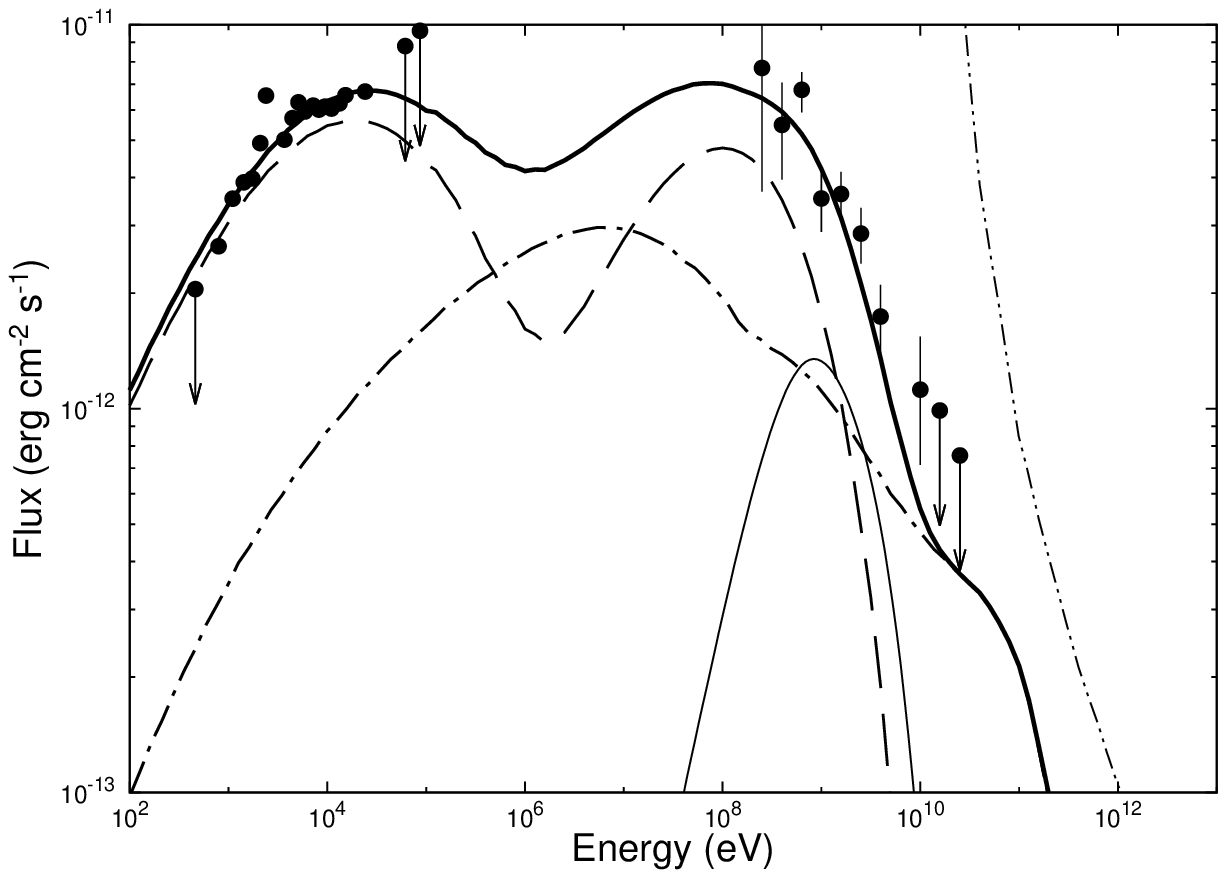}
  \caption{Multi-wavelength  spectrum of PSR~J0540-6919. The dashed line and dashed-dotted line
    represent the calculated spectra of the  low-energy secondary pairs that are
    created by the primary curvature photons
    and the high-energy secondary pairs that were produced
    by primary TeV photons via  the inverse-Compton process, respectively.
    The thin sold line is the spectrum of the residual curvature photons
    from the outer gap.  The results are for the inclination
    angle $\alpha=10^{\circ}$ and viewing angle $\zeta=80^{\circ}$ (or $100^{\circ}$)  and $j_{in}=j_{out}=10^{-2}$.
    The dashed-double-dotted line is the expected sensitivity of the Cherenkov Telescope Array (Acharya et al. 2013).
    The observed X-ray data were taken from de~Plaa et al. (2013). For Fermi-LAT  data,
    we read the data from Ackermann et al. (2015) and multiplied  0.75 for estimating the pulsed fluxed, which is
    suggested in their paper. }
  \label{J0540spe}
\end{figure}
\label{fit}
\begin{figure}
\centering
\includegraphics[width=0.9\textwidth]{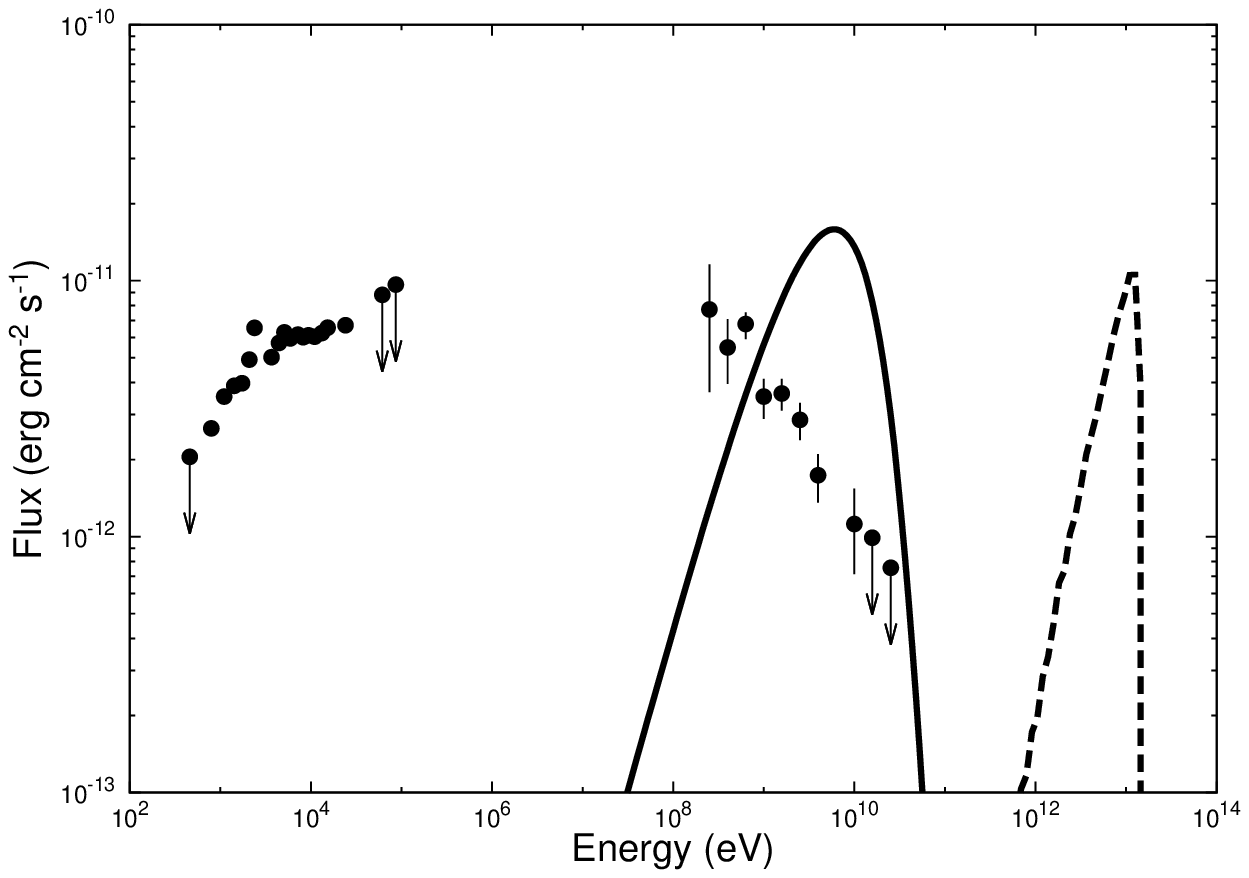}
\caption{Intrinsic gamma-ray fluxes (measured on the Earth) from the outer gap.
  The solid and dashed lines show the spectra of
the curvature radiation and the inverse-Compton scattering process, respectively.}
\label{intrinsic}
 \end{figure}
In this section, we apply the model to PSR~J0540-6919.  Since there are  two 
kinds of the secondary pairs in our calculation, we define 
the terminology ``low-energy secondary'' which represents
the pairs created by the primary curvature photons, and ''high-energy secondary'' for those  created
  by primary TeV photons via the inverse-Compton scattering process.

Figure~\ref{J0540spe} shows the multi-wavelength spectrum of PSR~J0540-6919 with the  model fitting curves. In the figure, 
the dashed line shows the synchrotron  and inverse-Compton scattering processes of the low-energy secondary pairs,
and dashed-dotted line is the emissions from
the high-energy secondary pairs. The results are for the inclination angle
$\alpha=10^{\circ}$ and the observer viewing angle $\zeta=80^{\circ}$
(or $100^{\circ}$). In addition, we assume that the injection rate at the
inner and outer  boundaries is 1\% of the Goldreich-Julian value, $j_{in}=j_{out}=10^{-2}$. Figure~\ref{intrinsic} shows
the intrinsic spectra for the curvature radiation (solid line) and inverse-Compton scattering process (dashed line) 
 inside the gap.

\begin{figure}
    \centering
    \includegraphics[width=0.9\textwidth]{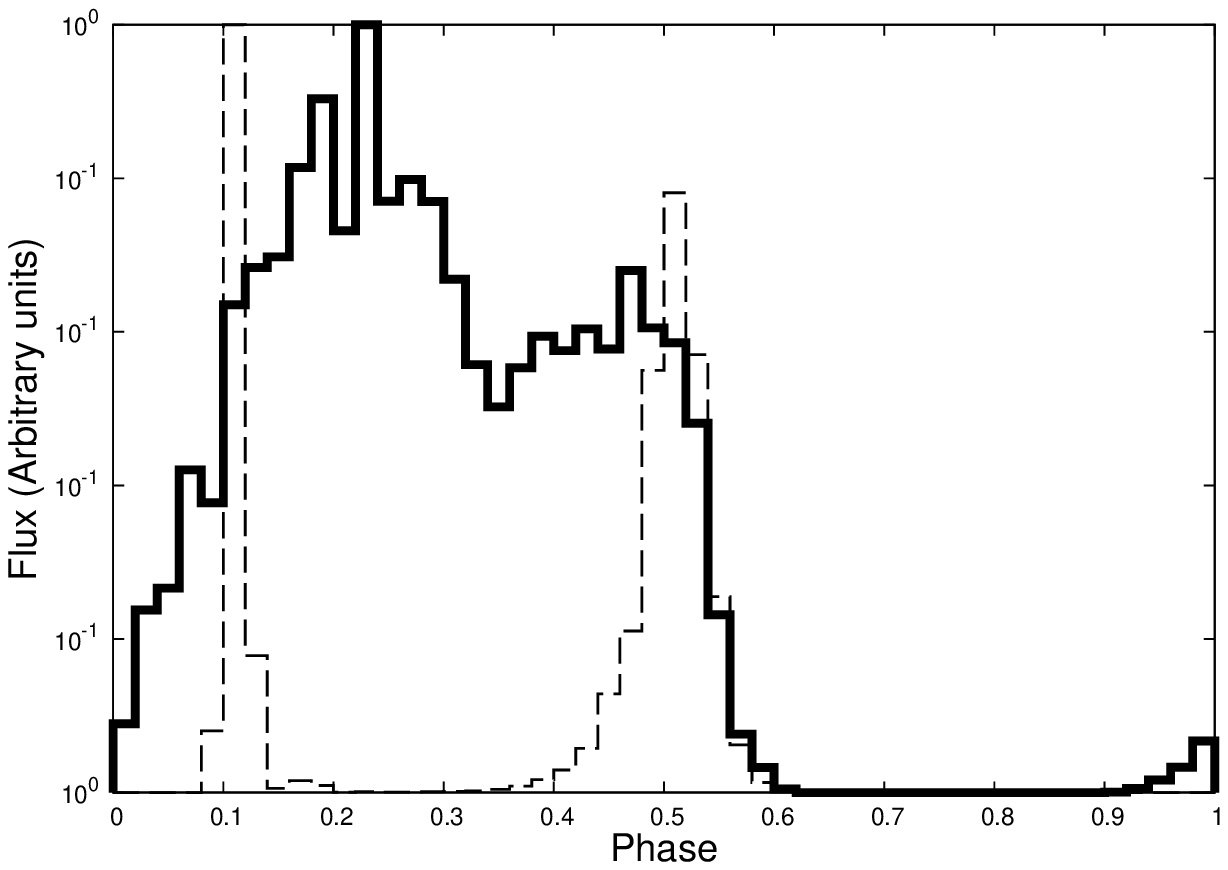}
    \caption{Calculated gamma-ray pulse profiles for the viewing angle $\zeta=80^{\circ}$. The inclination
    angle is $\alpha=10^{\circ}$ for solid line and $\alpha=70^{\circ}$ for dashed line, respectively.}
    \label{J0540light}
  \end{figure}

As we can see in Figure~\ref{J0540spe}, the emissions (dashed line) from
the low-energy secondary pairs explain the observed emissions in the 100eV-1GeV energy bands. However, the calculated spectrum above 1GeV  decays faster
than the Fermi-LAT data. To reconcile with Fermi-LAT data above 1GeV,  therefore,
the present model predicts that  the residual curvature emissions
(thin solid line) and/or the emissions from high-energy secondary pairs (dashed-dotted line) contribute to
the Fermi-LAT observations.  In the current calculation,
  a fraction of high-energy photons ($>$10GeV) emitted
  by the secondary pairs created near the light cylinder
  can escape from  the pair-creation process.
As we will argue in section~\ref{discuss}, the dynamic behavior of the  outer
gap discussed in  Takata et al. (2016) will not be the main reason to explain 
the observed spectrum above the cut-off energy  of PSR~J0540-6919.
With a small inclination angle $\alpha=10^{\circ}$, the calculated gamma-ray
  light curve (solid line in Figure~\ref{J0540light})
  shows a broad pulse with two narrow peaks separated
  by $\delta\phi\sim 0. 2$, which is consistent with the observations.

\subsection{Luminosity versus Inclination angle}
\begin{figure}
    \centering
    \includegraphics[width=0.9\textwidth]{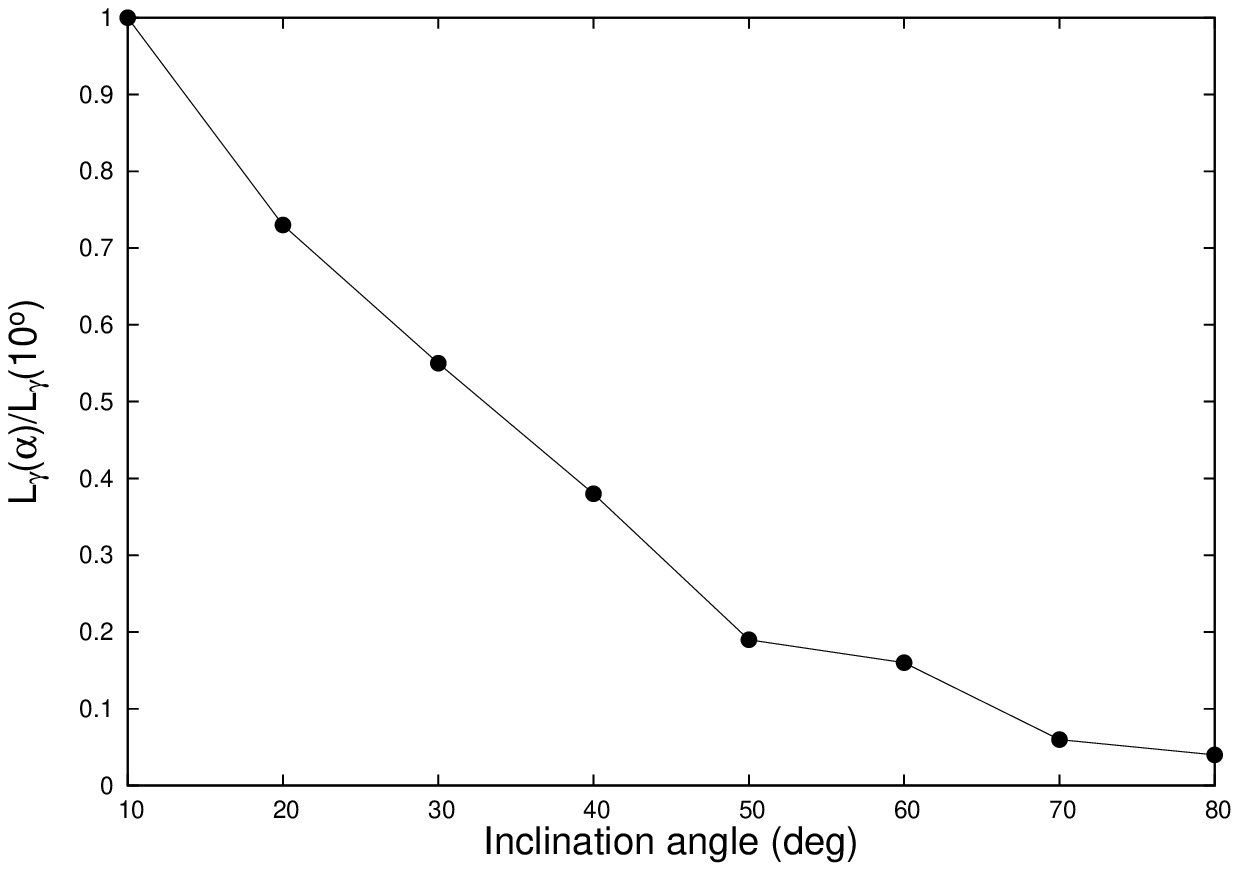}
    \caption{The calculated radiation luminosity with the function of the inclination angle. The vertical axis
      is normalized by the luminosity at $\alpha=10^{\circ}$. The results are
    for $j_{in}=j_{out}=10^{-2}$. }
    \label{lumi}
  \end{figure}

The Fermi-LAT  observations found that the efficiency, $\eta$, of  PSR~J0540-6919 
is about a factor of ten larger than the Crab pulsar, and this result
is incompatible with the empirical relation $\eta \propto L_{sd}^{-1/2}$ of the 
Fermi-LAT pulsars (Abdo et al. 2013). Here we suggest the
smaller magnetic inclination of PSR~J0540-6919 causes  the larger
radiation efficiency than the Crab pulsar,  whose
magnetic inclination angle will be  relatively large. 

Figure~\ref{lumi} shows the calculated gamma-ray luminosity
as a function of  the inclination angle.  In the figure,
the vertical axis is normalized by the calculated 
luminosity at $\alpha=10^{\circ}$. We find in the 
figure that  the calculation luminosity tends to decrease 
as the inclination angle increases.  In the current model, 
this dependency was caused by the dependency on (1) the 
position of the null charge surface of the Goldreich-Julian charge density and 
on (2)  the maximum gap current on the inclination angle.
The gap power depends  on the thickness of the gap in the poloidal plane, and  it  decreases with 
decreasing of the thickness. The electrodynamics of the conventional gap  models expects
the relation that $L_{\gamma}\sim  f_{gap}^{3}L_{sd}$,  where $f_{gap}$ is defined by the ratio between 
 the size of the outer gap measured on the stellar surface and  the polar cap size
 (Takata et al. 2010). For the Crab-like pulsars, 
the outer gap size may be controlled by
 the mean-free path of the pair-creation process
 between the curvature photons and soft X-rays from the neutron star surface (Wang et al. 2010).
 Since the null charge surface on the last-open 
field lines approaches to the stellar surface with the increase of the inclination angle, 
the location of the outer gap is closer to the stellar surface 
for larger inclination angle. Since the number density of
X-rays from the stellar surface is inversely proportional to 
 the square of the radial distance, the pair-creation mean-free path inside the gap
 is shorter for the outer gap closer to the stellar surface. Hence,
 the outer gap becomes thinner and as a result the gap
 radiation power decreases with the increasing of the inclination angle.

 In the current calculation, the pair-creation process inside the gap is occurring due to
 collision   between GeV gamma-rays and surface X-rays.
   In this case, the mean free path of the pair-creation process at  around  the light cylinder
   is estimated as $\lambda(R_{lc})\sim 100R_{lc}$. With this mean-free path and the 
   injection rate $j_{out}=0.01$, 
   the gap thickness is determined
   so as to produce $\sim 5\times 10^4$ curvature photons inside the outer gap by
 one particle injected at the outer boundary,  and thus to
 make $\sim 100$ of pairs by the pair-creation process inside
 the gap (see section~\ref{discuss}). With a constant mean-free path
 $\lambda=100_{lc}$, we find that the GeV photons have to travel
 a distance of $\sim 0.15R_{lc}$ inside the gap to screen the gap.
 The mean-free path actually depends on the position as $\lambda(r)\propto r^2$. For a smaller inclination angle,
 because the null charger surface is close
 to the light cylinder, a constant mean-free path with $\lambda(r)=\lambda_0$ is a good approximation.  For a larger inclination angle, on the other hand,
 the null charge surface is closer to the stellar surface
 and the radial dependency of the mean-free path becomes more important, indicating the average
 mean-free path is shorter. Therefore, the required
 travel distance of the gamma-rays to create $\sim 100$ pairs
 becomes shorter than $\sim 0.15R_{lc}$ of the lower inclination case, 
 and therefore the gap thickness reduces. 

 It has been suggested that the inner
 boundary of the middle part of the outer gap tends to be  shifted toward the stellar surface
as the gap current increases. The model suggests that  the inner boundary will touch on the stellar surface
if the gap current is $j_{gap}\sim \cos\alpha$ in units of the Goldreich-Julian value
(Takata et al. 2004), which decreases with increasing  inclination angle.
This is because the charge density that is created by the gap current at the inner boundary should
match with the local value of Goldreich-Julian charge density, which  on the polar cap region  is
$\sim \cos\alpha B_s/(P_sc)$. One may expect that if
the inner boundary of the outer gap once touches on the stellar surface,  the latter  supplies copious particles to close the
outer gap. Therefore, the gap luminosity decreases
with increasing  inclination angle.

As described in section~\ref{fit}, the  smaller magnetic inclination
angle of PSR~J0540-6919 preferentially explains the observed 
small separation of the two peaks in the pulse profile. For a larger
magnetic inclination angle $(\alpha\geq 50^{\circ})$ and a viewing
angle $\zeta\sim 90^{\circ}$,
the phase-separation between two peaks is $\delta\phi\sim 0.4-0.5$, as shown
in Figure~\ref{J0540light}, and this  would be the case for the Crab pulsar. 
We  emphasize, therefore, that the  smaller inclination magnetic
angle of PSR~J0540-6919 can explain both the higher 
radiation efficiency and the narrower
phase separations of the two peaks than those of the Crab pulsar.

\section{Discussion}
\label{discuss}
\subsection{PSR~J0537-6910}
Fermi-LAT resolved the gamma-ray emissions from the high spin-down  powered
pulsar, J0537-6910, in the  LMC with a flux level of
$F_{\gamma}\sim 10^{-11}{\rm erg~cm^{-2}~s^{-1}}$. However, the
pulsed emissions in Fermi-LAT data have yet to be confirmed, and the observed 
spectrum fitted by a single power-low function indicates the emissions  to be from the
pulsar wind nebula and/or a supernova remnant (Ackermann et al. 2015).
Since PSR~J0537-6910 has the largest spin-down power ($L_{sd}\sim 5\times 10^{38}~\rm{erg~s^{-1}}$) and the strongest magnetic
field at the light cylinder ($B_{lc}\sim 2\times 10^6$G) among the known pulsars (see the ATNF pulsar catalog, Manchester et al. 2005),
it is likely that this pulsar produces gamma-rays in the magnetosphere, 
and but they  are buried under the
background emission, or the gamma-ray beam is out of the line of  sight.
The observed emission properties of PSR~J0537-6910 are very different from those of
the Crab and J0537-6910; (1) the pulsed emissions have been discovered only in the X-ray bands, (2) the observed  radiation
efficiency in the X-rays  is very low $\eta_X\sim 3\times 10^{-4}$,
and  (3) the pulse width, $\sim 0.2$, in the  X-ray bands   (Marshall et al. 1998) is narrower than those of the other pulsars.

No detection of the pulsed emissions by the Fermi-LAT makes if  difficult for us
to discuss the electromagnetic spectrum  in the wide energy
bands, and  to constrain the   magnetic inclination and the Earth viewing angle.
However, we  may expect that the radiation process of PSR~J0537-6910
is similar to those of the Crab and J0540-6919, and we may assume that the flux level of the pulsed gamma-rays measured  on the Earth is 
$F_{\gamma}\sim F_{X}$, which is the case for the Crab and PSR~J0540-6919. Under those assumptions, the observed radiation
efficiency will be of the order of $\eta_{J0537}\sim 10^{-3}$, which is about two orders of  magnitude smaller than that of
J0540-6919. As expected from Figure~\ref{lumi}, we would say that it is difficult to explain $\eta_{J0540}/\eta_{J0537}\sim 100$
by the effect of the inclination angle. If both PSRs J0537-6910 and J0540-6919 have a viewing angle $\zeta\sim 90^{\circ}$,  
it is also difficult to explain the difference in the pulse  width with the difference in the inclination angle.
We suggest therefore that the Earth viewing angle  is very different between the two pulsars.
Our model  suggests that the Earth viewing angle  of PSR~J0540-6919 is
close to $\zeta\sim 90^{\circ}$ measured from the spin axis, which is also suggested by a  study of the pulsar wind
(Ng \& Romani 2004,2008).
Since most of the pairs inside the gap are created around the null charge surface,
the outer gap emission is stronger for an  Earth viewing angle of  $\zeta\sim 90^{\circ}$.
As the viewing angle deviates from the $\zeta\sim 90^{\circ}$, therefore, the observed gap emission rapidly decreases and hence
the apparent radiation efficiency decreases (see  figures~3 and 4 in  Takata et al. 2011); at the same time,
the pulse width becomes narrower.
On these grounds, we speculate  that the main reason for  difference in the observed efficiencies and in the observed pulsed widths between  PSRs~J0540-6919 and~J0537-6910
is the difference in the Earth viewing angle.

\subsection{Dependency on $j_{in}$ and $j_{out}$}
In Figure~\ref{J0540spe}, we assumed the same particle injection rates at the
inner and outer boundaries. The assumption of equal injection rates at
the gap boundaries is arbitrary, and it is not necessary for the real case. 
In the current {\it local} model, however, it would  not be possible to consistently solve  the injection
particles at the gap boundaries,  for which we would  have to solve the global structure including 
the polar cap activities, outer gap activities, and pulsar wind region.
To see the dependency on the choice of the injection current, we  examined 
the case for $j_{in}=0$ and   $j_{out}=0$, that is, no particles
enter into the gap from the inner boundary (star side) or outer boundary (light cylinder side), respectively.

\begin{figure}
  \centering
  \includegraphics[width=0.9\textwidth]{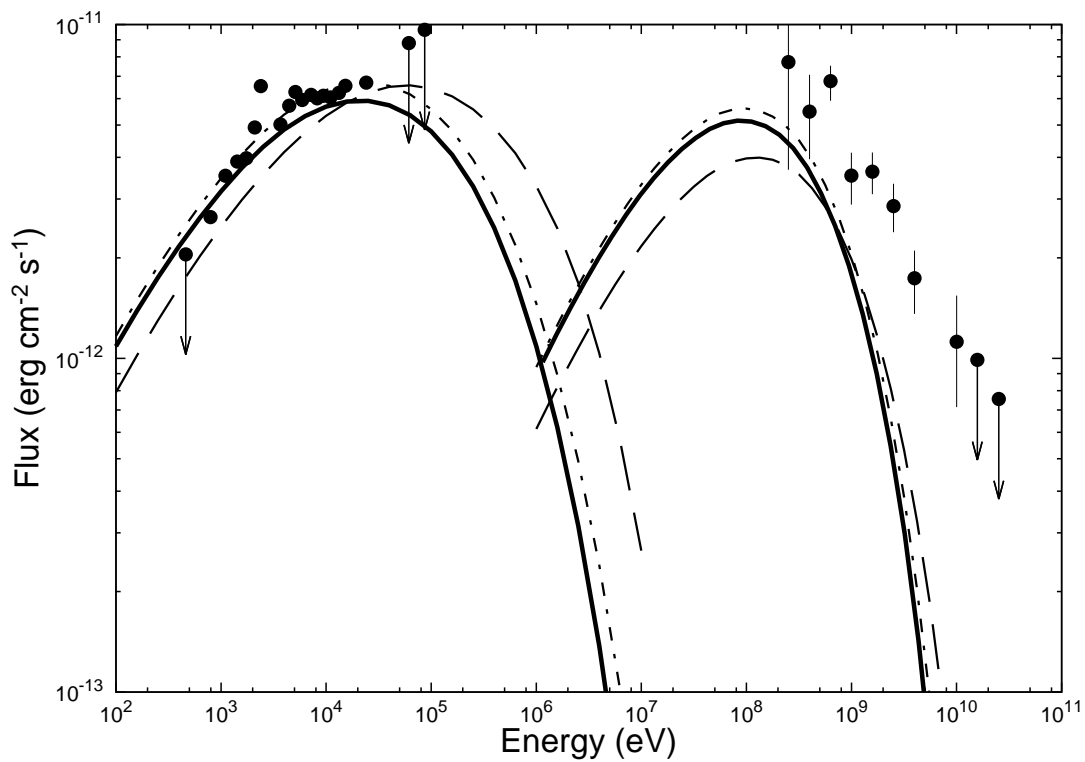}
  \caption{Multi-wavelength spectrum of PSR~J0540-6919. The lines show the spectra of synchrotron radiation(lower energy part)
    and inverse-Compton scattering process (higher energy part) of the low-energy secondary pairs.
    The sold lines, dashed lines and dashed-dotted lines are
    result for $(j_{in},j_{out})=(10^{-2}, 10^{-2})$, $(10^{-2},0)$ and $(0,10^{-2})$, respectively. In addition,
  we assume $\alpha=10^{\circ}$ and $\zeta=80^{\circ}$. }
  \label{J0540spe2}
\end{figure}

Figure~\ref{J0540spe2} summarizes the dependency of the emissions from the low-energy secondary pairs
on the injection currents $j_{in}$ and $j_{out}$; the solid line, dashed line
and dashed-dotted line are results for $(j_{in},j_{out})=(10^{-2},10^{-2})$, $(10^{-2},0)$ and  $(0,10^{-2})$, respectively. We
find in the figure that the calculated spectra become  harder for $j_{out}=0$ (dashed line in Figure~\ref{J0540spe2}). This is
related to the fact that most of the pairs are created by the inwardly propagating gamma-rays. Collision
with the  X-rays from the surface  is  a head-on process for inwardly propagating gamma-rays, while it is tail-on process for
outwardly propagating gamma-rays. Hence, the mean-free path of the former  is shorter than
that of latter, and most of the pairs are created by the inwardly propagating gamma-rays. This indicates
that the gap size is mainly controlled by the pair-creation process of
the inwardly propagating gamma-rays.  For $j_{out}=0$, therefore,
the outer gap has to be thick to create enough pairs,   and as a result
the calculated spectrum becomes harder.

\begin{figure}
\centering
\includegraphics[width=0.9\textwidth]{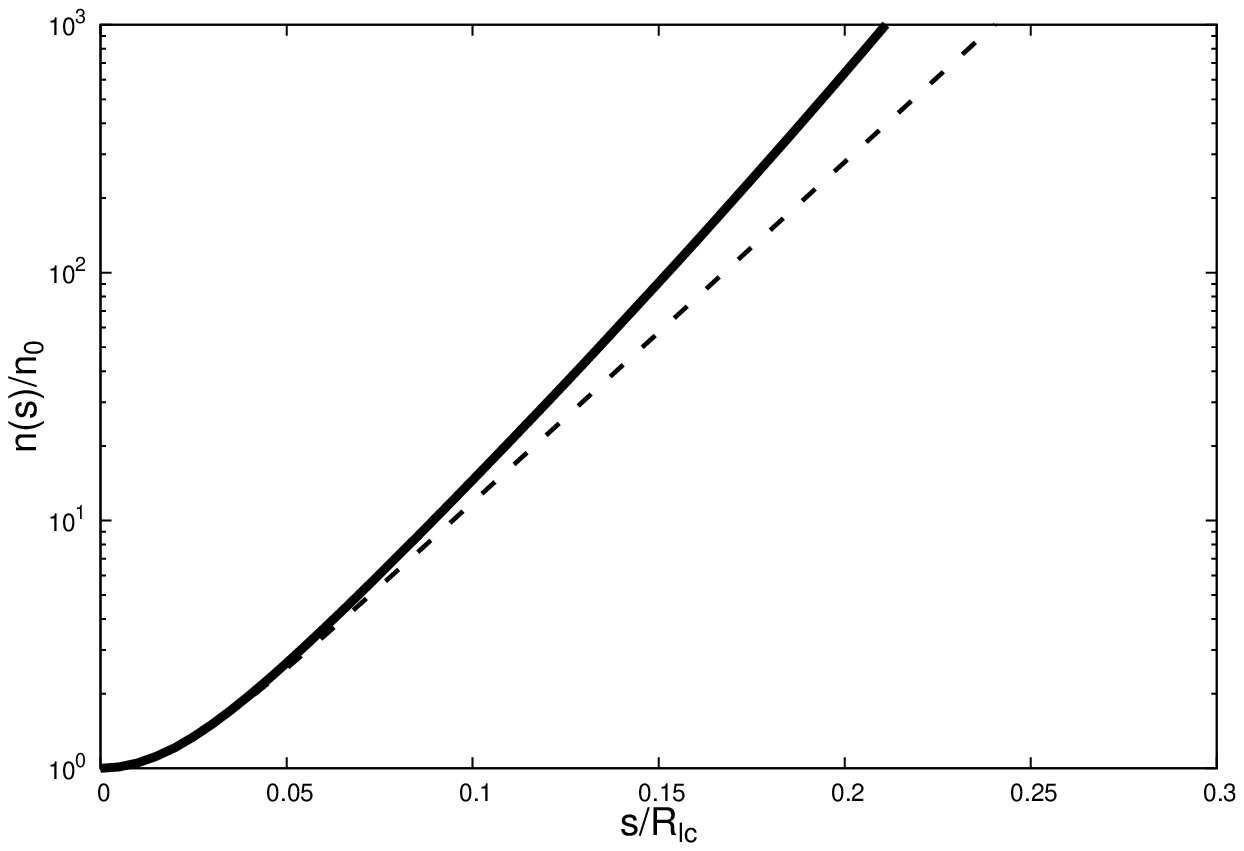}
\caption{Evolution of the number of the inwardly moving
  particles in the pair-creation regions along the magnetic field line from the outer boundary.
  The solid and dashed line show the solutions for
  the pair-creation mean free path of $\lambda\propto (1-s/R_{lc})^2$ (\ref{sol2})
 and $\lambda$=constant (\ref{sol1}), respectively.
}
\label{pair}
\end{figure}

  In our  calculation, the gap structure is controlled by the magnitude of
  $j_{out}$, except for the case $j_{out}\ll j_{in}$. We quantitatively discuss
  how the gap size depends on the injection rate $j_{out}$.
  Since the gap thickness of the Crab-like pulsar is about 10\% of the light cylinder
  radius, we can approximate that the propagation direction of the gamma-rays
  is the same as the direction of the particle's motion.
  Under this approximation,
  the evolution of the number density of the inwardly moving particles (electrons)
  and gamma-rays in the pair-creation region is  described by 
  \begin{equation}
    \frac{dn_-(s)}{ds}=\frac{g_-(s)}{\lambda(s)}, 
    \label{dn}
  \end{equation}
  and
  \begin{equation}
     \frac{dg_-(s)}{ds}=P_cn_-(s),
\label{dg}
  \end{equation}
  respectively, where $s$ is the distance from the outer boundary, and 
  $n_-$ and $g_-$ are number  density and photon number density normalized by the
  Goldreich-Julian density, respectively. We ignore the effect of the pair-creation by
  the gamma-rays propagating outward.
  In addition, $\lambda(s)$ and $P_c$ are  the
  mean free path of the photon-photon pair-creation process and the rate
  of the curvature radiation.  In the present calculation, since we assume
  the surface temperature $T_s\sim 10^6$K, the mean-free path inside the gap 
  at the light cylinder is estimated as
  $\lambda_0\sim 1/(\sigma_{\gamma\gamma}n_X)\sim 100R_{lc}$, where
  we used $\sigma_{\gamma\gamma}=0.2\sigma_{T}$ and $n_X\sim \sigma_{SB}R_{s}^2
  T^3/(ck_BR_{lc}^2)\sim 3\times 10^{14}{\rm cm^{-3}}$ with $\sigma_{SB}$ being
  Stefan-Boltzmann constant. The rate of the curvature radiation is estimated
  as $P_c\sim 3\times 10^4/R_{lc}(\Gamma/10^7)(R_c/R_{lc})^{-1}$, where $R_c$ being the curvature radius,
  and $\Gamma$ the Lorentz factor of the accelerated particles.

  To solve the equations~(\ref{dn}) and (\ref{dg}), we impose the boundary conditions
  as $n_-(0)=n_o$  and $g_-(0)=0$, where $s=0$ represents the outer boundary.
  We assume that the rate of the curvature radiation process, $P_c$,
  is constant along the magnetic field line.
  By assuming  that mean free path is constant along the distance $s$ from the outer boundary
  ($\lambda(r)= \lambda_0$),  we find  the solution that
  \begin{equation}
 n_-(s)=\frac{n_o}{2}\left({\rm e}^{c_1s}+{\rm e}^{-c_1s}\right),
\label{sol1}
  \end{equation}
  where $c_1=(P_c/\lambda_0)^{1/2}$.

  Since we consider the surface X-ray emission as
  the soft-photon field for the photon-photon pair-creation process inside
  the outer gap, the mean-free path will decrease as $\lambda\propto r^2$.
  To take into account this effect, we explore the solution with
  the mean-free path in the form  $\lambda(s)=\lambda_0(1-s/R_{lc})^2$. The solution
   becomes 
  \begin{equation}
    n_-(s)=\frac{n_o}{b}\left[a_+\left(1-\frac{s}{R_{lc}}\right)^{a_+-1}
      -a_-\left(1-\frac{s}{R_{lc}}\right)^{a_--1}\right],
    \label{sol2}
  \end{equation}
  where $a_{\pm}=(1\pm b)/2$ with $b=\sqrt{1+4P_cR_{lc}^2/\lambda_0}$.
  Figure~\ref{pair} shows the evolution of the ratio of the local number density
  and that  at the outer boundary of the inwardly moving particles,  $n(s)/n_o$,
  as a function of the distance from the outer boundary. The solid and dashed
lines are solutions given by equations~(\ref{sol2}) and~(\ref{sol1}),
respectively; here we adopted $P_c=10^5/R_{lc}$ and $\lambda_0=10^2R_{lc}$.

We can find in the figure that the multiplicity of the particles
injected at the outer boundary becomes  $\sim 100$ if the
gamma-rays travel $\sim 0.15R_{lc}$ from the outer boundary. This suggests that
if the injection rate at the boundary is 1\% of the Goldreich-Julian value
(that is, $j_{out}=0.01$), the number density  becomes the Goldreich-Julian value
after the gamma-ray travels  $\sim 0.15R_{lc}$ in the outer gap and the created pairs will significantly
screen the accelerating electric field. This is consistent with the
gap structure solved in this paper.  Figure~\ref{pair} also
indicates that for a smaller injection rate,
the gamma-rays have to travel a greater  distance to achieve  
the Goldreich-Julian number density of the pairs
by the pair-creation, and hence the gap size becomes larger.

As Figure~\ref{J0540spe} shows,  the current model with using constant injection rate  $(j_{in}, j_{out})$
predicts that  the residual curvature radiation of the primary particles
and/or the emissions from the high-energy secondary pairs can explain the observed
emissions above 1GeV.
Takata et al. (2016), on the other hand,  argued that sub-exponential decays of
the GeV spectra of the Fermi-LAT pulsars
reflect the time-dependent emission process  of the outer gap.
They proposed that  the injection rate at the gap boundaries is  time-dependent
variable and the observed gamma-ray spectrum is emitted from different gap structures with different injection rates.
In the model, the observed spectrum was fitted better  as
the superposition of several power-law plus exponential cut-off functions with
varying  the cut-off energy, for which
the different components are produced at the different injection rates at the gap boundaries.

\begin{figure}
  \centering
  \includegraphics[width=0.9\textwidth]{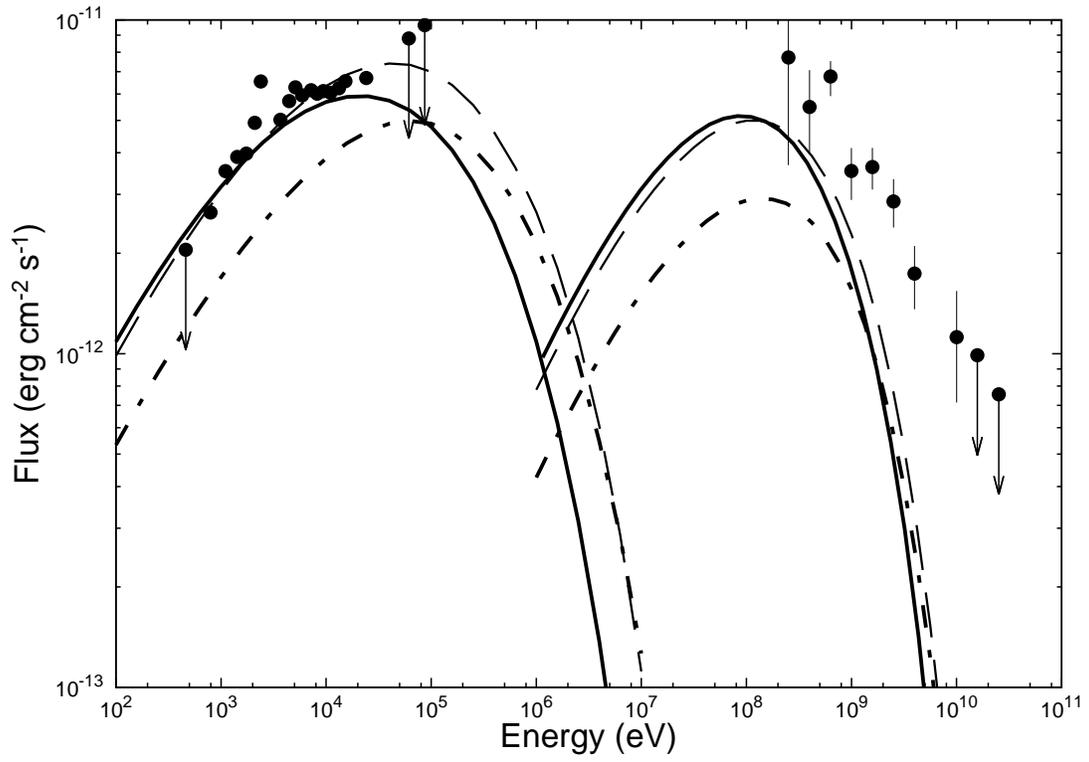}
  \caption{Same as Figure~\ref{J0540spe2}, but $(j_{in},j_{out})=(10^{-2}, 10^{-2})$ for solid lines, $(10^{-3},10^{-3})$ for dashed lines and $(10^{-4},10^{-4})$ for dashed-dotted lines. }
\label{J0540spe1}
\end{figure}

We  discuss  the shape of the observed GeV spectrum of PSR~J0540-6191 by the Fermi-LAT with the dynamics
model in  Takata et al. (2016).  We find however that the calculated GeV spectra do not greatly affect the  assumed
 extent of injection rate at the gap boundaries.  This is because the GeV gamma-rays observed on the
Earth  do not come from  the primary pairs in   the gap, but the  secondary pairs that are decelerated by the radiation process.
Figure~\ref{J0540spe1} shows the spectra of the emissions from the low-energy secondary pairs calculated  with
different injection rates; $j_{in}=j_{ou}=10^{-2}$  (solid line),
$10^{-3}$ (dashed line) and $10^{-4}$ (dotted line). We can see that the  hardness (peak energy) 
of the ``synchrotron emissions'' (low-energy component)  increases with decreasing of the injection rate. This is because the
gap thickness increases and hence the electric field in the gap becomes stronger as the injection rate decreases.
As a result, the energy distribution of the low-energy secondary pairs that emit synchrotron photons
becomes harder for a gap with a smaller injection rate.
On the other hand, the energy peak of the spectra by ``inverse-Compton scattering''
(high-energy component) does not greatly depend on the injection rate.  This is because the synchrotron cooling is more
important for the particles with a Lorentz  factor $\Gamma>200$ than the radiation
cooling,  due to the inverse-Compton scattering. As Figure~\ref{J0540spe1} shows,
therefore, the energy peak of
the inverse-Compton scattering of low-energy secondary pairs always appears around $\sim 100$MeV, regardless
of  the injection rates.  As a result,
it is obvious from Figure~\ref{J0540spe1} that even if we superpose the emissions calculated with
different injection rates, the combined spectrum decays faster and  still
has a large  discrepancy with the Fermi-LAT spectrum  above  1GeV.
On these grounds, we conclude
that the residual curvature photons and/or the emissions
from the high-energy secondary pairs contribute to the observed emissions
above 1GeV.

In summary, we discussed the gamma-ray emissions from the Crab-like pulsars, PSRs~J0537-6910 and~J0540-6919, in the LMC.
The pulsed emissions from  PSR~J0540-6919 is observed to have an  efficiency that is a factor of ten larger than that of
the Crab pulsar.  By solving the electrodynamics of the outer gap accelerator, we concluded that the difference in the
radiation efficiencies of  PSR~J0540-6919 and the Crab pulsar is caused  by the difference in the inclination angle.
Inferred from the observed X-ray emissions, the radiation efficiency of PSR~J0537-6910 is about two orders of magnitude smaller
than that of
PSR~J0540-6919. Because of the  very narrow X-ray pulse and low radiation efficiency of PSR~J0537-6910, we suspect that
the Earth viewing angle of PSR~J0537-6910 greatly deviates from $\zeta\sim 90^{\circ}$, which is the case for the Crab and PSR~J0540-6919.

The authors  thank  K. Hirotani, the referee, for insightful comments and suggestions on the manuscript.  We al thank
A. H. Kong, C. Y. Hui, P. H. T. Tam, M. Ruderman,
and S. Shibata for the useful discussions. JT is supported by
the NSFC grants of China under 11573010.
KSC is supported by a GRF grant of the Hong Kong Government
under HKU17300814P.
All calculations were done under  the High Performance Computing  Cluster (Hyperion) of the  Institute of Particle Physics
 and Astrophysics, HUST. 

\end{document}